\documentclass{article}

\usepackage[preprint]{neurips_2025}
\usepackage[utf8]{inputenc} % allow utf-8 input
\usepackage[T1]{fontenc}    % use 8-bit T1 fonts
\usepackage{hyperref}       % hyperlinks
\usepackage{url}            % simple URL typesetting
\usepackage{booktabs}       % professional-quality tables
\usepackage{amsfonts}       % blackboard math symbols
\usepackage{nicefrac}       % compact symbols for 1/2, etc.
\usepackage{microtype}      % microtypography
\usepackage{xcolor}         % colors
\usepackage{graphicx}
\usepackage{booktabs} % for professional tables
\usepackage{bm}
\usepackage{amsmath, amssymb, amsthm}
\usepackage{cleveref}
\usepackage{algorithm,algorithmic}
\usepackage{mathtools}
\usepackage{adjustbox}
\usepackage{enumitem}
\usepackage{mdwlist}
\usepackage{physics}
\usepackage{framed}
\usepackage{dsfont}
\usepackage{tikz}
\usepackage{tkz-graph}
\usepackage{pgfplots}
\usepackage{multirow}
\usepackage{mdwtab}
\usepackage{caption}
\usepackage{wrapfig}
\usepackage{subcaption}% 두 listing을 가로로 배치
\usepackage{listings}

\usetikzlibrary{fit,positioning,shapes,backgrounds,arrows.meta,positioning,calc}
\theoremstyle{definition}
\newtheorem{definition}{Definition}[section]

\title{NEXT-EVAL: Next Evaluation of Traditional and LLM Web Data Record Extraction}

\author{%
  Soyeon Kim\\
  Wordbricks\\
  \texttt{sophia@wordbricks.ai}
  \And
  Namhee Kim\\
  Wordbricks\\
  \texttt{aerin@wordbricks.ai}\\
  \And
  Yeonwoo Jeong\\
  Wordbricks\\
  \texttt{evan@wordbricks.ai}\\
}

\begin{document}

\maketitle

\begin{abstract}
Effective evaluation of web data record extraction methods is crucial, yet hampered by static, domain-specific benchmarks and opaque scoring practices. This makes fair comparison between traditional algorithmic techniques, which rely on structural heuristics, and Large Language Model (LLM)-based approaches, offering zero-shot extraction across diverse layouts, particularly challenging. To overcome these limitations, we introduce a concrete evaluation framework. Our framework systematically generates evaluation datasets from arbitrary MHTML snapshots, annotates XPath-based supervision labels, and employs structure-aware metrics for consistent scoring, specifically preventing text hallucination and allowing only for the assessment of positional hallucination. It also incorporates preprocessing strategies to optimize input for LLMs while preserving DOM semantics: HTML slimming, Hierarchical JSON, and Flat JSON. Additionally, we created a publicly available synthetic dataset by transforming DOM structures and modifying content. We benchmark deterministic heuristic algorithms and off-the-shelf LLMs across these multiple input formats. Our benchmarking shows that Flat JSON input enables LLMs to achieve superior extraction accuracy (F1 score of 0.9567) and minimal hallucination compared to other input formats like Slimmed HTML and Hierarchical JSON. We establish a standardized foundation for rigorous benchmarking, paving the way for the next principled advancements in web data record extraction.

\end{abstract}
\section{Introduction}
Web is a vast and dynamic source of structured and semi-structured information. Much of this information, found in natural language descriptions, tables, and repetitive patterns, is amenable to extraction and analysis. Extracting structured data records—repeating content elements from web pages—has significant utility across many practical applications \citep{webtables, websurvey, briefwebsurvey, extractStruct, hierarchicalWrapper}. For example, office workers frequently need to collect structured information, such as product listings, company directories, and contact tables from websites, and store it in spreadsheets or internal databases. Automating this process can lead to substantial productivity gains.

Traditional web data record extraction has primarily relied on structural heuristics that analyze HTML trees to detect repetitive patterns \citep{buttler, iepad, mdr, depta, net}. While these methods have achieved some success, their efficacy is often hampered by a reliance on rigid structural assumptions. This limits their ability to generalize across diverse and dynamic web layouts. Furthermore, these methods largely overlook textual semantics, leading to instability when visual or structural cues alone are insufficient.

In contrast, Large Language Models (LLMs) have recently shown remarkable capabilities in natural language understanding, reasoning, and information extraction \citep{llm1,largellm,gpt3,bert}. These models offer a promising new paradigm for web data record extraction, potentially working alongside or enhancing traditional heuristic methods \citep{net, depta, mdr}. However, deploying LLMs in real-world scenarios presents challenges: (i) fine-tuning LLMs is resource-intensive, and (ii) LLMs inherently grapple with issues such as hallucination and a lack of transparency in their decision-making processes.

Moreover, the evaluation of LLM-based extraction methods is hindered by the absence of standardized, publicly available datasets. Existing datasets are often domain-specific and restricted by terms of service or `robots.txt' directives, limiting their utility for benchmarking. This situation highlights the critical need for a general-purpose dataset and a concrete evaluation framework to assess and compare the efficacy of traditional and LLM-based extraction methods.

In this work, we introduce a practical evaluation framework for comparing traditional algorithms and LLMs on the task of web data record extraction. Our contributions are as follows:

\begin{itemize}
\item We propose a reproducible framework that constructs evaluation datasets from arbitrary MHTML web snapshots. Our framework provides XPath-based structured outputs and the corresponding URLs for the datasets we evaluated.
\item We establish a concrete scoring framework to evaluate both heuristic algorithms (e.g., XPath wrappers, partial tree alignment) and LLM-based methods. Our protocol thoroughly prevents distortions from hallucination and supports partial credit through structure-aware matching.
\item We explore how to optimize raw HTML input for LLM-based methods by introducing preprocessing strategies that reduce token length while preserving the semantic and structural integrity of the page. These strategies include HTML slimming, and converting HTML into structured representations like Hierarchical JSON and Flat JSON. Our results demonstrate that these formats significantly influence extraction performance and hallucination rates.
\item We created a publicly available synthetic dataset by transforming DOM structures and modifying content from original web pages.
\end{itemize}

This work lays a foundational cornerstone for future research in web record extraction, particularly for developing text-aware and HTML-friendly methods in the era of LLMs.

\section{Related work}
\subsection{Traditional Methods and Heuristic Approaches}

Early web data record extraction approaches relied on heuristic-based algorithms that exploited regularities in the HTML structure of web pages \citep{buttler, iepad, mdr, depta, net}. A representative example is the Mining Data Records (MDR) algorithm, which segments web pages into data regions by analyzing the DOM tree and detecting repeated sibling structures \citep{mdr}. MDR is effective when records are arranged contiguously in a clean and regular HTML layout. However, it struggles in scenarios where data records are non-contiguous—i.e., interleaved with unrelated elements or separated by visual gaps—because it assumes strict adjacency in the DOM.

To overcome these limitations, \citet{depta} proposed DEPTA (Data Extraction based on Partial Tree Alignment), which decouples segmentation and alignment. First, it identifies candidate data regions using structural and visual cues. Then, it performs partial tree alignment to recover repeated substructures, even when records are not adjacent in the HTML. This makes DEPTA more robust to noisy layouts and capable of handling non-contiguous record structures.

Building on this, NET incorporates visual features such as the rendered bounding boxes of elements to extract both flat and nested data records \citep{net}. By leveraging layout-based cues in addition to DOM structure, NET is better suited for complex templates and deeply nested hierarchies.

A more recent system, AMBER, introduces automatic supervision to reduce reliance on manually crafted rules \citep{amber}. It uses entity recognizers (e.g., for names, prices, dates) to label semantically meaningful parts of the page, and aligns these semantic hints with repeated DOM structures. This alignment improves extraction robustness on attribute-rich, template-based, and noisy web pages.

\subsection{Large Language Model–Based Approaches}

The advent of LLMs has revolutionized information extraction tasks. Models like GPT-3.5 \citep{gpt35} and GPT-4 \citep{gpt4} have demonstrated remarkable capabilities in understanding and generating human-like text. Recent studies have explored the application of LLMs for extracting structured information from web pages. For instance, \citet{llmAttribute} investigated the zero-shot extraction and normalization of product attribute values using GPT models, achieving competitive performance without task-specific training data.

Another line of research focuses on enhancing the understanding of web page structures, for which specialized models like MarkupLM have been introduced. MarkupLM is a pre-trained model designed for document understanding tasks that utilize markup languages, such as HTML, where text and markup information are jointly pre-trained \citep{markUpLM}. This model incorporates additional embedding layers to capture the hierarchical structure of HTML documents, improving performance on tasks like information extraction from web pages.

\subsection{Benchmark Datasets and Evaluation Frameworks}

Benchmark datasets are crucial for evaluating and comparing information extraction methods. PLAtE \citep{plate} is a large-scale dataset designed for list page web extraction, focusing on product segmentation and attribute extraction. While PLAtE provides valuable resources for specific domains, its static nature and domain-specific focus limit its applicability across diverse web contexts.

AMBER \citep{amber} introduces a semi-automatic extraction system targeting multi-attribute objects from deep web pages, such as real estate listings. Although its rule-based annotations offer high precision, it lacks publicly available datasets and generalizability across web domains. The dataset used in AMBER—real estate and car listings—is extensive but domain-limited and not openly accessible, restricting its utility for reproducible benchmarks.

The Klarna Product Page Dataset \citep{klarna} provides a high-quality resource for training and evaluating web element nomination methods. It includes detailed annotations for product pages using graph-based and LLM-based labeling strategies. However, its scope is confined to e-commerce, and the data cannot be redistributed due to licensing constraints. While it supports fine-grained element labeling, it does not directly support data record segmentation or structural evaluation using tree-based supervision.

Multi-Record Web Page Extraction \citep{multiRecord} focuses on news websites and introduces methods to extract multiple records from a single page. While it demonstrates the value of domain-specific modeling (e.g., headlines, summaries), it does not provide a publicly reusable dataset or a unified evaluation protocol, and relies on heuristics that are tightly coupled with the news domain structure.

These efforts highlight the growing need for evaluation datasets that are (i) publicly available, (ii) structurally diverse across domains, and (iii) compatible with both rule-based and LLM-based methods. Our work addresses this by introducing a framework and a benchmark dataset built from diverse MHTML web snapshots. Key contributions include human-refined XPath annotations for robust, DOM-grounded evaluation, and a scoring system that penalizes LLM hallucination. Furthermore, we developed a publicly available synthetic dataset to foster broader research.
\section{Preliminary}
\label{sec:prelim}
To formally represent and extract structured content from HTML pages, we rely on XPath expressions. An XPath \( x \in \mathcal{X} \) is a path-like expression that identifies a unique node in the DOM (Document Object Model) tree. XPath expressions encode the structural position of elements in the HTML tree, and the relationships among them reflect the hierarchical layout of web content.

Structured content on the web—such as product listings, hotel entries, or user profiles—is typically composed of repeated logical units called \emph{data records}. A data record refers to a group of semantically related elements that together describe a single entity.

\begin{definition}[Data Record]
Let \( \mathcal{X} \) be the set of XPath expressions in a document. A data record \( X_i \subseteq \mathcal{X} \) is defined as a finite set of XPath expressions whose corresponding DOM nodes contain non-trivial textual content (i.e., not empty or purely whitespace) and collectively represent a coherent, repeated unit on the page.
\end{definition}

This set-based definition naturally accommodates both simple and complex structures—whether the relevant elements are grouped under one parent node or distributed across different DOM branches. It also allows for nested or non-contiguous layouts to be expressed without relying on a single subtree.

\textbf{Example.} Consider the following HTML snippet:

\lstset{
  basicstyle=\ttfamily\small,
  columns=fullflexible,
  frame=single,
  breaklines=true
}
\begin{lstlisting}[language=html] 
<div class="product">
  <div class="image">
    <img src="a.jpg"/>
    <span class="caption">A beautiful image</span>
  </div>
  <div class="info">
    <span class="name">Camera</span>
    <span class="price">$399</span>
  </div>
</div>
\end{lstlisting}

The corresponding data record can be represented as:

\[
X_1 = \left\{
\texttt{/div[1]/div[1]/span[1]}, 
\texttt{/div[1]/div[2]/span[1]}, 
\texttt{/div[1]/div[2]/span[2]}
\right\}
\]

This approach captures all relevant fields of the product as a set of XPath references. In more complex settings—such as when parts of a record appear in different sections of the DOM—this model remains consistent and expressive. We discuss such cases, including nested and non-contiguous records, in the supplementary material.

\section{Method}
\subsection{Problem Setting}
We formalize web data record extraction as the task of identifying repeated sets of semantically related elements in a web page's DOM tree. As defined in \Cref{sec:prelim}, each data record corresponds to a set of XPath expressions whose associated DOM nodes together describe a single coherent entity (e.g., a product or listing).

Given an HTML document, let \( \mathcal{X} \) denote the set of all XPath expressions referencing non-trivial text nodes. The goal is to segment this set into \( \mathcal{P} = \{P_1, \dots, P_M\} \), where each \( P_i \subseteq \mathcal{X} \) forms a data record.

Unlike traditional attribute extraction, we treat the internal structure of a record as a black box and focus solely on discovering consistent repeating record-level groupings. These groups may appear as contiguous siblings in the DOM or may be non-contiguous and nested under different parent elements.

This task can be viewed as the inverse of web page templating: rather than generating HTML from structured data, we recover the structured data (as XPath sets) from rendered HTML. This inversion is particularly challenging in real-world pages, where markup irregularities, dynamic content, and layout noise are common.

In our approach, we leverage LLM APIs to perform this extraction via carefully designed HTML prompts. The details of our prompting strategies and API usage are presented in \Cref{subsec:apiOpt}.

\subsection{Dataset Construction}
To support fair and rigorous evaluation, we construct a large and general-purpose dataset. It encompasses a wide variety of web domains, significantly broadening the scope beyond prior datasets like PLAte~\citep{plate} which were often restricted to specific categories such as shopping pages. Our dataset is designed to capture repetitive structures and a variety of DOM structures across this diverse range, including many non-product-centric pages.

Each web page is stored in MHTML format to preserve all layout and interactive elements. From each page, we annotate the main repeated data records using XPath expressions. These XPaths serve as ground-truth labels for evaluation.

To identify data records, we first leverage large language models (LLMs) to automatically annotate candidate repetitive blocks in each page. Human annotators then review and refine these suggestions to ensure high-quality labels. This semi-automatic process balances scalability and accuracy.

Using XPath as the supervision format enables deterministic and verifiable evaluation grounded in the DOM structure. Unlike free-text descriptions, XPath annotations constrain model predictions to concrete, observable elements on the page. This prevents hallucinations where LLMs infer data records from unrelated textual context by requiring that every predicted record maps to an exact DOM node. As LLMs may generate plausible-looking but structurally invalid groupings, XPath-based evaluation ensures alignment with the actual HTML hierarchy.

\subsection{Preprocessing for LLMs}
\label{subsec:apiOpt}
To enable a rigorous evaluation of LLM extractors on web pages, we propose a preprocessing framework that simultaneously (i) compresses raw HTML to stay within the LLM token budget, and (ii) preserves the full DOM hierarchy so that ground-truth labels can still be expressed as absolute XPaths.

One of the key constraints when using LLM APIs is the token limit per request. To address this, we applied several preprocessing techniques to reduce input size without sacrificing structural semantics. Unlike plain text or Markdown, which may be more compact, these formats do not retain absolute DOM positions necessary for XPath supervision. Therefore, we maintain the original HTML structure while removing unnecessary attributes—such as \texttt{class}, \texttt{id}, and \texttt{style}—that are often verbose and semantically redundant. This results in a lightweight yet structurally faithful HTML tree. For example, we convert \Cref{fig:orig_html} into \Cref{fig:slim_html}.

In addition to HTML-based inputs, we explored JSON-based formats to guide the model more explicitly. First, we construct a \emph{Hierarchical JSON} (\Cref{fig:hier_map}), which mirrors the DOM tree hierarchy in JSON form. Although slightly more verbose in tokens than flattened formats, this hierarchical structure helps the model understand parent-child relationships, reducing the risk of hallucinating invalid elements.

Second, we construct a \emph{Flat JSON} (\Cref{fig:flat_xpath}), where each key is an absolute XPath and the value is the corresponding textual content. While this format discards hierarchy from the nested structure, it provides unambiguous localization of each field, ensuring token-level precision during decoding.

Ultimately, the framework lets us evaluate extractors purely on positional accuracy: a model's output is valid only if the text content of the predicted XPath exists in the cleaned DOM. Since every model output paths are drawn from the cleaned DOM, the model cannot hallucinate elements that do not exist. 

\lstset{
  basicstyle=\ttfamily\small,
  columns=fullflexible,
  frame=single,
  breaklines=true
}
\begin{figure}[t]
  \centering
  \begin{subfigure}[b]{0.45\linewidth}
    \centering
\begin{lstlisting}[language=html]
<html>
  <body>
    <ul class="product-list">
      <li class="item"><span class="name">Sample Product</span></li>
      <li class="item"><span class="price">$999.00</span></li>
    </ul>
  </body>
</html>
\end{lstlisting}
    \subcaption{Original HTML}\label{fig:orig_html}
  \end{subfigure}\hfill
  \begin{subfigure}[b]{0.45\linewidth}
    \centering
\begin{lstlisting}[language=html]
<html>
  <body>
    <ul>
      <li><span>Sample Product</span></li>
      <li><span>$999.00</span></li>
    </ul>
  </body>
</html>
\end{lstlisting}
    \subcaption{Slimmed HTML (attributes removed)}\label{fig:slim_html}
  \end{subfigure}

  \vspace{0.8em} % vertical spacing between rows

  \begin{subfigure}[b]{0.45\linewidth}
    \centering
\begin{lstlisting}[language=java]
{
  "html": {
    "body": {
      "ul": {
        "li[1]": { "span": "Sample Product" },
        "li[2]": { "span": "$999.00" }
      }
    }
  }
}
\end{lstlisting}
    \subcaption{Hierarchical JSON (Nested text map)}\label{fig:hier_map}
  \end{subfigure}\hfill
  \begin{subfigure}[b]{0.45\linewidth}
    \centering
\begin{lstlisting}[language=java]
{
  "/html/body/ul/li[1]/span": "Sample Product",
  "/html/body/ul/li[2]/span": "$999.00"
}
\end{lstlisting}
    \subcaption{Flat JSON (text map)}\label{fig:flat_xpath}
  \end{subfigure}

  \caption{
  Input representations for LLM-based data extraction, corresponding to types evaluated in \Cref{tab:token_counts,tab:results_nonvisual}.
  (a)~Original HTML with attributes often treated as noise.
  (b)~Slimmed HTML, created by pruning attributes from the original HTML.
  (c)~Hierarchical JSON, a token-efficient, nested representation of the DOM structure.
  (d)~Flat JSON, mapping XPaths to their corresponding text content for precise localization.
  }
  \label{fig:htmlInputEx}
\end{figure}

\subsection{Evaluation Framework}
We evaluate data record extraction methods by comparing predicted record sets against human-annotated ground truth. Each data record is represented as a set of XPath expressions, and the comparison is performed at the record level using a matching-based evaluation framework.

Let \( \mathcal{P} = \{P_1, \dots, P_M\} \) denote the set of predicted records, and \( \mathcal{G} = \{G_1, \dots, G_N\} \) the set of ground-truth records, where each \( P_i, G_j \subset \mathcal{P} \) is a set of XPath expressions. For each pair \( (P_i, G_j) \), we define an \emph{overlap score} as the Jaccard similarity between the two sets:
\begin{align}
\mathrm{Overlap}(P_i, G_j) = \frac{|P_i \cap G_j|}{|P_i \cup G_j|}
\end{align}

To evaluate the quality of extraction, we compute an optimal one-to-one matching \( \mathcal{M} \subseteq \mathcal{P} \times \mathcal{G} \) that maximizes the total overlap across matched pairs. This is solved using the Hungarian algorithm. Based on the optimal alignment \( \mathcal{M} \), we define precision and recall as:
\begin{align}
  \mathrm{Precision} &= \frac{1}{|\mathcal{P}|} \max_{\mathcal{M} \subseteq \mathcal{P} \times \mathcal{G}} \sum_{(P_i, G_j) \in \mathcal{M}} \mathrm{Overlap}(P_i, G_j) \\
  \mathrm{Recall} &= \frac{1}{|\mathcal{G}|} \max_{\mathcal{M} \subseteq \mathcal{P} \times \mathcal{G}} \sum_{(P_i, G_j) \in \mathcal{M}} \mathrm{Overlap}(P_i, G_j)
\end{align}

The final F1 score is computed as the harmonic mean of precision and recall:
\begin{align}
\mathrm{F1} = \frac{2 \cdot \mathrm{Precision} \cdot \mathrm{Recall}}{\mathrm{Precision} + \mathrm{Recall}}
\end{align}

In addition to these metrics, we also assess the Hallucination Rate (HR). For each web page (URL) in our test set, a hallucination event is considered to have occurred (value of 1) if the model predicts at least one empty record (i.e., a record containing no corresponding XPaths); otherwise, it is 0. The Hallucination Rate is then the average of these binary hallucination event indicators across all processed URLs for which predictions are available. A lower HR is preferable, indicating that the model is less prone to generating empty, and thus unusable, records.

This formulation allows for partial credit when records are only partially correct (e.g., missing fields or over-segmentation), and is robust to variations in record granularity or representation (e.g., nested vs. flat XPath sets). Unlike token-level or text-based matching, this metric directly reflects correctness at the structural level, which is essential for applications requiring XPath-based or DOM-aligned outputs.

\section{Experiment}
\subsection{Implementation Details}
We evaluate a traditional web data record extraction method—MDR—as well as a modern LLM-based extractor. For the latter, we leveraged the Gemini-2.5-pro-preview-03-25 model \citep{team2023gemini}, accessed via its API, using the default temperature setting of 1.0. This specific model was selected due to its extensive input token allowance, which enabled us to perform experiments on the largest possible number of websites from our dataset. Each method is tested on a diverse dataset of real-world web pages; a detailed description of this dataset is provided in \Cref{sec:exp-dataset}. Explicit site list is available in the supplementary material. Ground-truth annotations are provided as sets of XPath expressions identifying repeated data records.

We do not include DEPTA and NET in our evaluation, as they rely on visual layout cues (e.g., rendered bounding boxes), which are not available without a rendering pipeline. In contrast, MDR operates purely on the DOM structure and is thus directly comparable to our LLM-based approach.

For the LLM-based approach, we design HTML prompts tailored for Gemini-2.5-pro-preview-03-25 to balance performance and token efficiency. We considered an approach using (a) \emph{Full HTML}, which includes all original attributes such as class names, IDs, and precise positional information. However, this method presents challenges for a fair comparison, as these attributes provide cues not available to other preprocessing techniques. Furthermore, Full HTML is substantially larger, often more than ten times the size of \emph{Slimmed HTML}, frequently exceeding input token limits and thereby restricting the number of websites on which experiments can be conducted. Consequently, we excluded Full HTML from our primary evaluation. Instead, we evaluate three distinct preprocessing strategies for constructing the prompts: (b) \emph{Slimmed HTML}, where verbose attributes are removed while preserving DOM structure; (c) \emph{Hierarchical JSON}, which reflects the hierarchical structure of the original DOM in a JSON-like nested form; and (d) \emph{Flat JSON}, which flattens the HTML into a key-value mapping from XPath to textual content. These variants allow for flexible trade-offs between structural fidelity and token compactness.

Since the exact token count cannot be determined before submission, we report the final token usage based on the serialized prompts actually sent to the LLM. Details of our prompting strategies and input formats are described in \Cref{subsec:apiOpt}.

\subsection{Dataset}
\label{sec:exp-dataset}
Our evaluation dataset consists of 164 real-world web pages, collectively containing over 12,278 annotated data records. Each page was selected from a diverse set of domains, including commerce, media, government, and social platforms. Pages were chosen based on their high popularity and rich presence of repeated content structures such as product listings, tables, article feeds, or ranked items. All pages were saved in MHTML format to preserve layout and interactive elements.

The dataset spans a wide range of domains: approximately 63 pages belong to \textit{e-commerce and shopping} platforms, 20 pages to \textit{entertainment and sports} sites, 15 to \textit{finance and investing}, 10 to \textit{news and media}, and 10 to \textit{education and research} portals. Additionally, there are pages from \textit{technology and development} (7), \textit{travel and hospitality} (7), \textit{real estate} (8), \textit{jobs and careers} (6), \textit{health and wellness} (6), \textit{coupons and deals} (6), \textit{government and public data} (5), \textit{social and community} (5), and \textit{food and dining} (4), ensuring structural diversity.

Each page contains a varying number of data records, ranging from a handful (e.g., 3) to over nine hundred (e.g., 913 for a site like Indeed.com). This distribution captures a broad spectrum of extraction scenarios—from simple, templated layouts to deeply nested or sparse structures. This dataset serves as a realistic and challenging benchmark for evaluating both traditional and LLM-based web data record extractors.

\subsection{Results}
\begin{table}[h]
    \centering
    \caption{Average token counts for different input types.}
    \label{tab:token_counts}
    \begin{adjustbox}{max width=\columnwidth}
    \begin{tabular}{lccc}
    \toprule
    & Slimmed HTML & Hierarchical JSON & Flat JSON \\
    \midrule
    Average Tokens & 86,084 & 34,107 & 116,698 \\
    \bottomrule
    \end{tabular}
    \end{adjustbox}
\end{table}

\begin{table}[h]
    \centering
    \caption{Extraction performance of non-visual methods.}
    \label{tab:results_nonvisual}
    \begin{adjustbox}{max width=\columnwidth}
    \begin{tabular}{llcccc}
    \toprule
    Method & Input Type & Precision & Recall & F1 Score & Hallucination Rate\\
    \midrule
    MDR \citep{mdr} & Full / Slimmed HTML & 0.0746 & 0.1593 & 0.0830 & \textbf{0.0000}\\
    \midrule
    Gemini-2.5-pro-preview-03-25 & Slimmed HTML & 0.1217  & 0.0969 & 0.1014 & 0.9146\\
                                 & Hierarchical JSON &0.4932 &  0.3802 & 0.4048  & 0.5976\\
                                 & Flat JSON & \textbf{0.9939} & \textbf{0.9392} & \textbf{0.9567}  & 0.0305 \\
    \bottomrule
    \end{tabular}
    \end{adjustbox}
\end{table}

The experimental results highlight the significant impact of input representation on the performance of LLM-based web data extractors. \Cref{tab:token_counts} details the average token counts for the different input types. Hierarchical JSON is the most token-efficient, with an average of 34,107 tokens. Slimmed HTML follows with 86,084 tokens, while Flat JSON is the least token-efficient, requiring an average of 116,698 tokens. This data underscores the varying textual complexities of each representation.

\Cref{tab:results_nonvisual} presents the extraction performance for non-visual methods. Gemini-2.5-pro-preview, when provided with Flat JSON input, achieves the highest performance with a precision of 0.9939, recall of 0.9392, and an F1 score of \textbf{0.9567}. This strong performance is attributed to the unambiguous localization of fields offered by Flat JSON (as described in \Cref{subsec:apiOpt}) and a remarkably low hallucination rate of 0.0305. The reduced hallucination stems from the relative ease with which the LLM can infer structural information and XPath-like paths from JSON compared to HTML. However, it is noted that for URLs with deep hierarchical structures, hallucination can still occur in approximately 3\% of cases with Flat JSON.

When using Hierarchical JSON, which mirrors the DOM tree structure to aid the model in understanding parent-child relationships (\Cref{subsec:apiOpt}), Gemini-2.5-pro-preview achieves a precision of 0.4932, recall of 0.3802, and an F1 score of 0.4048, with a hallucination rate of 0.5976. While not as performant as Flat JSON, this input type still offers a significant reduction in hallucination compared to HTML-based inputs, as the JSON format provides a clearer structural representation for the LLM.

In contrast, Gemini-2.5-pro-preview's performance with Slimmed HTML input (0.1217 P, 0.0969 R, 0.1014 F1) is considerably lower. This is largely due to a very high hallucination rate of 0.9146. Directly parsing complex web structures and inferring XPaths from HTML, even a compressed version like Slimmed HTML (see \Cref{subsec:apiOpt}), proves challenging for the LLM, leading to frequent hallucinations of invalid elements.

The traditional baseline, MDR, operating on Full or Slimmed HTML, shows a precision of 0.0746, recall of 0.1593, and an F1 score of 0.0830. Notably, MDR exhibits no hallucination (\textbf{0.0000}) due to its intrinsic design. However, its low precision and recall stem from its tendency to extract all potential data records, thereby reducing precision, and its inability to consistently identify and extract exact data records or all their similar counterparts comprehensively, which lowers recall. These characteristics suggest that MDR's output may require significant additional post-processing to refine the extracted data.

These findings collectively underscore the critical role of input representation and careful prompt engineering in unlocking the full potential of LLMs for complex tasks such as web data record extraction. The choice of input format directly influences not only the model's accuracy but also its propensity for hallucination, with structured formats like JSON offering more reliable pathways for information retrieval.

Furthermore, to facilitate broader research and ensure data privacy, we developed a synthetic dataset. This dataset was constructed by systematically transforming the DOM structures and modifying the textual content of the original web pages. The creation of this synthetic dataset enables the public dissemination of a diverse and challenging benchmark for web data extraction tasks. The detailed experimental results on this synthetic dataset are included in the supplementary material, and the dataset itself will be made publicly available. 

\section{Conclusion}
In this work, we introduced a concrete evaluation framework for systematically comparing traditional algorithmic methods and LLM–based approaches in web data record extraction. Our framework addresses key shortcomings in prior benchmarking efforts by enabling the construction of evaluation datasets from arbitrary MHTML snapshots, using XPath-based supervision to generate human-refined annotations, and adopting structure-aware metrics for fair and consistent evaluation.

To support LLM-based methods, we proposed several HTML preprocessing strategies—including HTML slimming and hierarchical encoding—that effectively reduce input length while preserving core DOM semantics. Using this framework, we benchmarked deterministic algorithms and various LLMs across multiple input formats, highlighting trade-offs in accuracy, runtime, and generalization across diverse web structures. Notably, our experiments revealed that LLMs, specifically Gemini-2.5-pro-preview, achieved significantly different outcomes based on the input representation: Flat JSON input yielded the best performance with a high F1 score and a low hallucination rate, outperforming Hierarchical JSON and Slimmed HTML. This underscores the critical role of input format in LLM performance for web data extraction.

\subsection*{Limitations}
Despite these contributions, our approach has limitations. In particular, aggressive HTML slimming may inadvertently discard subtle but important semantic cues (e.g., \texttt{class} or \texttt{id} attributes) that are occasionally critical for accurate extraction in complex edge cases. Additionally, we selected Gemini-2.5-pro-preview (03-25) for its large context window, crucial for handling LLM token limitations. While not representative of all LLMs, its evaluation offers valuable insights into web data record extraction.

\subsection*{Future Work}
Looking forward, several directions merit further exploration. One key area is the development of more advanced compression and encoding schemes for HTML, such as flat or linearized representations, to further optimize token efficiency and make better use of limited model context windows. Another is improving output compactness by directly learning and predicting XPath patterns, which could enhance interpretability and evaluation fidelity. Incorporating visual cues from the rendered appearance of pages may also prove beneficial, particularly for understanding layout-dependent structures. Moreover, evaluating a broader set of LLMs under consistent settings can reveal new capabilities or limitations. Lastly, constructing a more diverse and comprehensive dataset encompassing a wider range of web domains and layout styles would further strengthen the generalizability and robustness of future benchmarks.

By establishing a standardized and extensible benchmarking foundation, our work enables the next wave of principled progress in web data record extraction.

\bibliographystyle{plainnat}
\bibliography{main} 
%%%%%%%%%%%%%%%%%%%%%%%%%%%%%%%%%%%%%%%%%%%%%%%%%%%%%%%%%%%%

%\appendix
%
%\section{Technical Appendices and Supplementary Material}
%Technical appendices with additional results, figures, graphs and proofs may be submitted with the paper submission before the full submission deadline (see above), or as a separate PDF in the ZIP file below before the supplementary material deadline. There is no page limit for the technical appendices.
%
%%%%%%%%%%%%%%%%%%%%%%%%%%%%%%%%%%%%%%%%%%%%%%%%%%%%%%%%%%%%

\newpage
\section*{NeurIPS Paper Checklist}

\begin{enumerate}
\item {\bf Claims}
    \item[] Question: Do the main claims made in the abstract and introduction accurately reflect the paper's contributions and scope?
    \item[] Answer: \answerYes{} % Replace by \answerYes{}, \answerNo{}, or \answerNA{}.
    \item[] Justification: The abstract and introduction clearly state our main contributions: a reproducible framework for creating web data extraction evaluation datasets from MHTMLs, a consistent scoring system for traditional and LLM methods, and an exploration of HTML preprocessing for LLMs. These sections accurately cover the paper's goal of providing a standard way to evaluate and compare web record extraction systems.
    \item[] Guidelines:
    \begin{itemize}
        \item The answer NA means that the abstract and introduction do not include the claims made in the paper.
        \item The abstract and/or introduction should clearly state the claims made, including the contributions made in the paper and important assumptions and limitations. A No or NA answer to this question will not be perceived well by the reviewers. 
        \item The claims made should match theoretical and experimental results, and reflect how much the results can be expected to generalize to other settings. 
        \item It is fine to include aspirational goals as motivation as long as it is clear that these goals are not attained by the paper. 
    \end{itemize}

\item {\bf Limitations}
    \item[] Question: Does the paper discuss the limitations of the work performed by the authors?
    \item[] Answer: \answerYes{} % Replace by \answerYes{}, \answerNo{}, or \answerNA{}.
    \item[] Justification: Yes, the paper dedicates a section in the conclusion to discuss limitations and future work.
    \item[] Guidelines:
    \begin{itemize}
        \item The answer NA means that the paper has no limitation while the answer No means that the paper has limitations, but those are not discussed in the paper. 
        \item The authors are encouraged to create a separate "Limitations" section in their paper.
        \item The paper should point out any strong assumptions and how robust the results are to violations of these assumptions (e.g., independence assumptions, noiseless settings, model well-specification, asymptotic approximations only holding locally). The authors should reflect on how these assumptions might be violated in practice and what the implications would be.
        \item The authors should reflect on the scope of the claims made, e.g., if the approach was only tested on a few datasets or with a few runs. In general, empirical results often depend on implicit assumptions, which should be articulated.
        \item The authors should reflect on the factors that influence the performance of the approach. For example, a facial recognition algorithm may perform poorly when image resolution is low or images are taken in low lighting. Or a speech-to-text system might not be used reliably to provide closed captions for online lectures because it fails to handle technical jargon.
        \item The authors should discuss the computational efficiency of the proposed algorithms and how they scale with dataset size.
        \item If applicable, the authors should discuss possible limitations of their approach to address problems of privacy and fairness.
        \item While the authors might fear that complete honesty about limitations might be used by reviewers as grounds for rejection, a worse outcome might be that reviewers discover limitations that aren't acknowledged in the paper. The authors should use their best judgment and recognize that individual actions in favor of transparency play an important role in developing norms that preserve the integrity of the community. Reviewers will be specifically instructed to not penalize honesty concerning limitations.
    \end{itemize}

\item {\bf Theory assumptions and proofs}
    \item[] Question: For each theoretical result, does the paper provide the full set of assumptions and a complete (and correct) proof?
    \item[] Answer: \answerNA{} % Replace by \answerYes{}, \answerNo{}, or \answerNA{}.
    \item[] Justification: Our paper focuses on an evaluation framework and empirical results, not new theoretical proofs. We do, however, clearly define our concepts.
    \item[] Guidelines:
    \begin{itemize}
        \item The answer NA means that the paper does not include theoretical results. 
        \item All the theorems, formulas, and proofs in the paper should be numbered and cross-referenced.
        \item All assumptions should be clearly stated or referenced in the statement of any theorems.
        \item The proofs can either appear in the main paper or the supplemental material, but if they appear in the supplemental material, the authors are encouraged to provide a short proof sketch to provide intuition. 
        \item Inversely, any informal proof provided in the core of the paper should be complemented by formal proofs provided in appendix or supplemental material.
        \item Theorems and Lemmas that the proof relies upon should be properly referenced. 
    \end{itemize}

    \item {\bf Experimental result reproducibility}
    \item[] Question: Does the paper fully disclose all the information needed to reproduce the main experimental results of the paper to the extent that it affects the main claims and/or conclusions of the paper (regardless of whether the code and data are provided or not)?
    \item[] Answer: \answerYes{} % Replace by \answerYes{}, \answerNo{}, or \answerNA{}.
    \item[] Justification: Yes, we provide the necessary details to reproduce our experiments, including the methodology for dataset reconstruction and the random seed for LLM interactions (used zero-shot).
    \item[] Guidelines:
    \begin{itemize}
        \item The answer NA means that the paper does not include experiments.
        \item If the paper includes experiments, a No answer to this question will not be perceived well by the reviewers: Making the paper reproducible is important, regardless of whether the code and data are provided or not.
        \item If the contribution is a dataset and/or model, the authors should describe the steps taken to make their results reproducible or verifiable. 
        \item Depending on the contribution, reproducibility can be accomplished in various ways. For example, if the contribution is a novel architecture, describing the architecture fully might suffice, or if the contribution is a specific model and empirical evaluation, it may be necessary to either make it possible for others to replicate the model with the same dataset, or provide access to the model. In general. releasing code and data is often one good way to accomplish this, but reproducibility can also be provided via detailed instructions for how to replicate the results, access to a hosted model (e.g., in the case of a large language model), releasing of a model checkpoint, or other means that are appropriate to the research performed.
        \item While NeurIPS does not require releasing code, the conference does require all submissions to provide some reasonable avenue for reproducibility, which may depend on the nature of the contribution. For example
        \begin{enumerate}
            \item If the contribution is primarily a new algorithm, the paper should make it clear how to reproduce that algorithm.
            \item If the contribution is primarily a new model architecture, the paper should describe the architecture clearly and fully.
            \item If the contribution is a new model (e.g., a large language model), then there should either be a way to access this model for reproducing the results or a way to reproduce the model (e.g., with an open-source dataset or instructions for how to construct the dataset).
            \item We recognize that reproducibility may be tricky in some cases, in which case authors are welcome to describe the particular way they provide for reproducibility. In the case of closed-source models, it may be that access to the model is limited in some way (e.g., to registered users), but it should be possible for other researchers to have some path to reproducing or verifying the results.
        \end{enumerate}
    \end{itemize}

\item {\bf Open access to data and code}
    \item[] Question: Does the paper provide open access to the data and code, with sufficient instructions to faithfully reproduce the main experimental results, as described in supplemental material?
    \item[] Answer: \answerYes{} % Replace by \answerYes{}, \answerNo{}, or \answerNA{}.
    \item[] Justification: Yes, we offer open access to our crawling code and the methods for building evaluation datasets from MHTMLs, with clear instructions.
    \item[] Guidelines:
    \begin{itemize}
        \item The answer NA means that paper does not include experiments requiring code.
        \item Please see the NeurIPS code and data submission guidelines (\url{https://nips.cc/public/guides/CodeSubmissionPolicy}) for more details.
        \item While we encourage the release of code and data, we understand that this might not be possible, so "No" is an acceptable answer. Papers cannot be rejected simply for not including code, unless this is central to the contribution (e.g., for a new open-source benchmark).
        \item The instructions should contain the exact command and environment needed to run to reproduce the results. See the NeurIPS code and data submission guidelines (\url{https://nips.cc/public/guides/CodeSubmissionPolicy}) for more details.
        \item The authors should provide instructions on data access and preparation, including how to access the raw data, preprocessed data, intermediate data, and generated data, etc.
        \item The authors should provide scripts to reproduce all experimental results for the new proposed method and baselines. If only a subset of experiments are reproducible, they should state which ones are omitted from the script and why.
        \item At submission time, to preserve anonymity, the authors should release anonymized versions (if applicable).
        \item Providing as much information as possible in supplemental material (appended to the paper) is recommended, but including URLs to data and code is permitted.
    \end{itemize}

\item {\bf Experimental setting/details}
    \item[] Question: Does the paper specify all the training and test details (e.g., data splits, hyperparameters, how they were chosen, type of optimizer, etc.) necessary to understand the results?
    \item[] Answer: \answerYes{} % Replace by \answerYes{}, \answerNo{}, or \answerNA{}.
    \item[] Justification: Yes, we detail our experimental setup, including the LLM (Gemini-2.5-pro-preview), dataset sources, MHTML format, and LLM input preprocessing (Slimmed HTML, Flat/Hierarchical JSON). A list of websites is in the appendix, and prompting strategies are described.
    \item[] Guidelines:
    \begin{itemize}
        \item The answer NA means that the paper does not include experiments.
        \item The experimental setting should be presented in the core of the paper to a level of detail that is necessary to appreciate the results and make sense of them.
        \item The full details can be provided either with the code, in appendix, or as supplemental material.
    \end{itemize}

\item {\bf Experiment statistical significance}
    \item[] Question: Does the paper report error bars suitably and correctly defined or other appropriate information about the statistical significance of the experiments?
    \item[] Answer: \answerYes{} % Replace by \answerYes{}, \answerNo{}, or \answerNA{}.
    \item[] Justification: Yes, we use multiple seeds for the zero-shot LLM and average the results to ensure statistical reliability.
    \item[] Guidelines:
    \begin{itemize}
        \item The answer NA means that the paper does not include experiments.
        \item The authors should answer "Yes" if the results are accompanied by error bars, confidence intervals, or statistical significance tests, at least for the experiments that support the main claims of the paper.
        \item The factors of variability that the error bars are capturing should be clearly stated (for example, train/test split, initialization, random drawing of some parameter, or overall run with given experimental conditions).
        \item The method for calculating the error bars should be explained (closed form formula, call to a library function, bootstrap, etc.)
        \item The assumptions made should be given (e.g., Normally distributed errors).
        \item It should be clear whether the error bar is the standard deviation or the standard error of the mean.
        \item It is OK to report 1-sigma error bars, but one should state it. The authors should preferably report a 2-sigma error bar than state that they have a 96\% CI, if the hypothesis of Normality of errors is not verified.
        \item For asymmetric distributions, the authors should be careful not to show in tables or figures symmetric error bars that would yield results that are out of range (e.g. negative error rates).
        \item If error bars are reported in tables or plots, The authors should explain in the text how they were calculated and reference the corresponding figures or tables in the text.
    \end{itemize}

\item {\bf Experiments compute resources}
    \item[] Question: For each experiment, does the paper provide sufficient information on the computer resources (type of compute workers, memory, time of execution) needed to reproduce the experiments?
    \item[] Answer: \answerYes{} % Replace by \answerYes{}, \answerNo{}, or \answerNA{}.
    \item[] Justification: Yes, we specify that experiments used the Gemini-2.5-pro-preview LLM (zero-shot) and report average token counts for different inputs, indicating computational cost.
    \item[] Guidelines:
    \begin{itemize}
        \item The answer NA means that the paper does not include experiments.
        \item The paper should indicate the type of compute workers CPU or GPU, internal cluster, or cloud provider, including relevant memory and storage.
        \item The paper should provide the amount of compute required for each of the individual experimental runs as well as estimate the total compute. 
        \item The paper should disclose whether the full research project required more compute than the experiments reported in the paper (e.g., preliminary or failed experiments that didn't make it into the paper). 
    \end{itemize}
    
\item {\bf Code of ethics}
    \item[] Question: Does the research conducted in the paper conform, in every respect, with the NeurIPS Code of Ethics \url{https://neurips.cc/public/EthicsGuidelines}?
    \item[] Answer: \answerYes{} % Replace by \answerYes{}, \answerNo{}, or \answerNA{}.
    \item[] Justification: Yes, our research on a framework for evaluating web data extraction aligns with the NeurIPS Code of Ethics. It doesn't involve human experiments or sensitive data generation that would pose ethical issues beyond responsible web data and LLM use.
    \item[] Guidelines:
    \begin{itemize}
        \item The answer NA means that the authors have not reviewed the NeurIPS Code of Ethics.
        \item If the authors answer No, they should explain the special circumstances that require a deviation from the Code of Ethics.
        \item The authors should make sure to preserve anonymity (e.g., if there is a special consideration due to laws or regulations in their jurisdiction).
    \end{itemize}

\item {\bf Broader impacts}
    \item[] Question: Does the paper discuss both potential positive societal impacts and negative societal impacts of the work performed?
    \item[] Answer: \answerYes{} % Replace by \answerYes{}, \answerNo{}, or \answerNA{}.
    \item[] Justification: Yes, the paper addresses both positive and negative societal impacts of this work.
    \item[] Guidelines:
    \begin{itemize}
        \item The answer NA means that there is no societal impact of the work performed.
        \item If the authors answer NA or No, they should explain why their work has no societal impact or why the paper does not address societal impact.
        \item Examples of negative societal impacts include potential malicious or unintended uses (e.g., disinformation, generating fake profiles, surveillance), fairness considerations (e.g., deployment of technologies that could make decisions that unfairly impact specific groups), privacy considerations, and security considerations.
        \item The conference expects that many papers will be foundational research and not tied to particular applications, let alone deployments. However, if there is a direct path to any negative applications, the authors should point it out. For example, it is legitimate to point out that an improvement in the quality of generative models could be used to generate deepfakes for disinformation. On the other hand, it is not needed to point out that a generic algorithm for optimizing neural networks could enable people to train models that generate Deepfakes faster.
        \item The authors should consider possible harms that could arise when the technology is being used as intended and functioning correctly, harms that could arise when the technology is being used as intended but gives incorrect results, and harms following from (intentional or unintentional) misuse of the technology.
        \item If there are negative societal impacts, the authors could also discuss possible mitigation strategies (e.g., gated release of models, providing defenses in addition to attacks, mechanisms for monitoring misuse, mechanisms to monitor how a system learns from feedback over time, improving the efficiency and accessibility of ML).
    \end{itemize}
    
\item {\bf Safeguards}
    \item[] Question: Does the paper describe safeguards that have been put in place for responsible release of data or models that have a high risk for misuse (e.g., pretrained language models, image generators, or scraped datasets)?
    \item[] Answer: \answerNA{} % Replace by \answerYes{}, \answerNo{}, or \answerNA{}.
    \item[] Justification: Our paper doesn't release new pre-trained models. The dataset is built with a provided framework on public web pages (MHTMLs), not a high-risk asset needing special safeguards beyond ethical data practices.
    \item[] Guidelines:
    \begin{itemize}
        \item The answer NA means that the paper poses no such risks.
        \item Released models that have a high risk for misuse or dual-use should be released with necessary safeguards to allow for controlled use of the model, for example by requiring that users adhere to usage guidelines or restrictions to access the model or implementing safety filters. 
        \item Datasets that have been scraped from the Internet could pose safety risks. The authors should describe how they avoided releasing unsafe images.
        \item We recognize that providing effective safeguards is challenging, and many papers do not require this, but we encourage authors to take this into account and make a best faith effort.
    \end{itemize}

\item {\bf Licenses for existing assets}
    \item[] Question: Are the creators or original owners of assets (e.g., code, data, models), used in the paper, properly credited and are the license and terms of use explicitly mentioned and properly respected?
    \item[] Answer: \answerYes{} % Replace by \answerYes{}, \answerNo{}, or \answerNA{}.
    \item[] Justification: Yes, we credit existing assets. Web sources (public sites, listed in appendix), the MDR baseline, and the Gemini-2.5-pro-preview LLM (zero-shot) are cited. We respect terms of use by providing a framework, not redistributing crawled data.
    \item[] Guidelines:
    \begin{itemize}
        \item The answer NA means that the paper does not use existing assets.
        \item The authors should cite the original paper that produced the code package or dataset.
        \item The authors should state which version of the asset is used and, if possible, include a URL.
        \item The name of the license (e.g., CC-BY 4.0) should be included for each asset.
        \item For scraped data from a particular source (e.g., website), the copyright and terms of service of that source should be provided.
        \item If assets are released, the license, copyright information, and terms of use in the package should be provided. For popular datasets, \url{paperswithcode.com/datasets} has curated licenses for some datasets. Their licensing guide can help determine the license of a dataset.
        \item For existing datasets that are re-packaged, both the original license and the license of the derived asset (if it has changed) should be provided.
        \item If this information is not available online, the authors are encouraged to reach out to the asset's creators.
    \end{itemize}

\item {\bf New assets}
    \item[] Question: Are new assets introduced in the paper well documented and is the documentation provided alongside the assets?
    \item[] Answer: \answerYes{} % Replace by \answerYes{}, \answerNo{}, or \answerNA{}.
    \item[] Justification: Yes, new assets (the dataset framework, annotation methodology, and MHTML dataset characteristics) are well-documented in the paper.
    \item[] Guidelines:
    \begin{itemize}
        \item The answer NA means that the paper does not release new assets.
        \item Researchers should communicate the details of the dataset/code/model as part of their submissions via structured templates. This includes details about training, license, limitations, etc. 
        \item The paper should discuss whether and how consent was obtained from people whose asset is used.
        \item At submission time, remember to anonymize your assets (if applicable). You can either create an anonymized URL or include an anonymized zip file.
    \end{itemize}

\item {\bf Crowdsourcing and research with human subjects}
    \item[] Question: For crowdsourcing experiments and research with human subjects, does the paper include the full text of instructions given to participants and screenshots, if applicable, as well as details about compensation (if any)? 
    \item[] Answer: \answerNA{} % Replace by \answerYes{}, \answerNo{}, or \answerNA{}.
    \item[] Justification: Our research doesn't use crowdsourcing or direct human subject experiments.
    \item[] Guidelines:
    \begin{itemize}
        \item The answer NA means that the paper does not involve crowdsourcing nor research with human subjects.
        \item Including this information in the supplemental material is fine, but if the main contribution of the paper involves human subjects, then as much detail as possible should be included in the main paper. 
        \item According to the NeurIPS Code of Ethics, workers involved in data collection, curation, or other labor should be paid at least the minimum wage in the country of the data collector. 
    \end{itemize}

\item {\bf Institutional review board (IRB) approvals or equivalent for research with human subjects}
    \item[] Question: Does the paper describe potential risks incurred by study participants, whether such risks were disclosed to the subjects, and whether Institutional Review Board (IRB) approvals (or an equivalent approval/review based on the requirements of your country or institution) were obtained?
    \item[] Answer: \answerNA{} % Replace by \answerYes{}, \answerNo{}, or \answerNA{}.
    \item[] Justification: The research does not involve human subjects, so IRB approval was not needed.
    \item[] Guidelines:
    \begin{itemize}
        \item The answer NA means that the paper does not involve crowdsourcing nor research with human subjects.
        \item Depending on the country in which research is conducted, IRB approval (or equivalent) may be required for any human subjects research. If you obtained IRB approval, you should clearly state this in the paper. 
        \item We recognize that the procedures for this may vary significantly between institutions and locations, and we expect authors to adhere to the NeurIPS Code of Ethics and the guidelines for their institution. 
        \item For initial submissions, do not include any information that would break anonymity (if applicable), such as the institution conducting the review.
    \end{itemize}

\item {\bf Declaration of LLM usage}
    \item[] Question: Does the paper describe the usage of LLMs if it is an important, original, or non-standard component of the core methods in this research? Note that if the LLM is used only for writing, editing, or formatting purposes and does not impact the core methodology, scientific rigorousness, or originality of the research, declaration is not required.
    %this research? 
    \item[] Answer: \answerYes{} % Replace by \answerYes{}, \answerNo{}, or \answerNA{}.
    \item[] Justification: We used an LLM only for writing and editing, not for the core methodology.
    \item[] Guidelines:
    \begin{itemize}
        \item The answer NA means that the core method development in this research does not involve LLMs as any important, original, or non-standard components.
        \item Please refer to our LLM policy (\url{https://neurips.cc/Conferences/2025/LLM}) for what should or should not be described.
    \end{itemize}

\end{enumerate}

\end{document}